\documentclass[12pt,longbibliography]{revtex4-1}

\usepackage{graphicx}
\usepackage{amsmath}
\usepackage{dcolumn}
\usepackage{enumitem}
\usepackage{epstopdf}

\begin{document}

\title{Ultra-strong photon-to-magnon coupling in multilayered heterostructures involving superconducting coherence via ferromagnetic layers}

\author{I.~A.~Golovchanskiy$^{1,2}$, N.~N.~Abramov$^{2}$, V.~S.~Stolyarov$^{1,3}$, M.~Weides$^{4}$, V.~V.~Ryazanov$^{2,5}$, A.~A.~Golubov$^{1,6}$, A.~V.~Ustinov$^{2,7}$, M.~Yu.~Kupriyanov$^{1,8}$}

\affiliation{
$^{1}$ Moscow Institute of Physics and Technology, State University, 9 Institutskiy per., Dolgoprudny, Moscow Region, 141700, Russia; \\
$^{2}$ National University of Science and Technology MISIS, 4 Leninsky prosp., Moscow, 119049, Russia; \\
$^{3}$ Dukhov Research Institute of Automatics (VNIIA), Sushchevskaya 22, Moscow 127055, Russia; \\
$^{4}$ James Watt School of Engineering, Electronics \& Nanoscale Engineering Division, University of Glasgow, Glasgow G12 8QQ, United Kingdom; \\
$^{5}$ Institute of Solid State Physics (ISSP RAS), Chernogolovka, 142432, Moscow region, Russia; \\
$^{6}$ Faculty of Science and Technology \& MESA+ Institute for Nanotechnology, University of Twente, 7500 AE Enschede, The Netherlands; \\
$^{7}$ Physikalisches Institut, Karlsruhe Institute of Technology, 76131 Karlsruhe, Germany; \\
$^{8}$ Skobeltsyn Institute of Nuclear Physics, MSU, Moscow, 119991, Russia
}%

\begin{abstract}
\newpage

\subsection*{Abstract}

The critical step for future quantum industry demands realization of efficient information exchange between different-platform hybrid systems, including photonic and magnonic systems, that can harvest advantages of distinct platforms. 
The major restraining factor for the progress in certain hybrid systems is the fundamentally weak coupling parameter between the elemental particles.
This restriction impedes the entire field of hybrid magnonics by making
realization of scalable on-chip hybrid magnonic systems unattainable.
In this work, we propose a general flexible approach for realization of on-chip hybrid magnonic systems with unprecedentedly strong coupling parameters.
The approach is based on multilayered micro-structures containing superconducting, insulating and ferromagnetic layers with modified both photon phase velocities and magnon eigen-frequencies.
Phenomenologically, the enhanced coupling strength is provided by the radically reduced photon mode volume.
The microscopic mechanism of the phonon-to-magnon coupling in studied systems evidences formation of the long-range superconducting coherence via thick strong ferromagnetic layers.
This coherence is manifested by coherent superconducting screening of microwave fields by the superconductor/ferromagnet/superconductor three-layers in presence of magnetization precession. 
This discovery offers new opportunities in microwave superconducting spintronics for quantum technologies.
\end{abstract}

\maketitle

\subsection*{Introduction}

The last decade has seen a remarkable progress in experimental quantum information sciences and in development of various artificial quantum systems.
Originally, experimental quantum physics was pioneered by the quantum optics, which introduced us to basic concepts of quantum information processing \cite{Flamini_RPP_82_016001}. 
With invention of superconducting qubits quantum technologies started to evolve with solid-state microwave superconducting quantum circuits \cite{Gu_PR_718_1}, whipping up the quantum computer race.

The next critical step in quantum industry demands realization of efficient information exchange between different-platform quantum systems \cite{Clerk_NatPhys_16_257,Xiang_RMP_85_623} that can harvest advantages of distinct systems. 
A number of perspective systems is based on coherent interaction of photons with mechanical oscillations (phonons) \cite{Aspelmeyer_RMP_86_391} or collective spin excitations (magnons) \cite{Lachance-Quirion_Sci_367_425} in ferromagnetic media.
The later is of particular interest for application in hybrid magnonic-based quantum platforms \cite{Tabuchi_CRR_17_729,Tabuchi_Sci_349_405,Lachance-Quirion_APE_12_070101,Wang_arXiv} and offers opportunities for development of novel quantum technologies such as magnon memory \cite{Zhang_NatComm_6_8914} or microwave-to-optical quantum transducers \cite{Hisatomi_PRB_93_174427}.

The major restraining factor for the progress in hybrid quantum magnonics is the fundamentally weak coupling parameter between photons and magnons.
This restriction can be circumvented by either a radical increase of a number of spins in a hybrid system, as done in cavity magnonics \cite{Tabuchi_PRL_113_083603, Zhang_PRL_113_156401, Lachance-Quirion_APE_12_070101,Flower_NJP_21_095004,Boventer_PRRes_2_013154}, or by engineering of on-chip microwave circuits with large local microwave fields \cite{Li_PRL_123_107701,Hou_PRL_123_107702}.
With both approaches the realization of scalable magnonic hybrids with freely-adjustable coupling characteristics remains unattainable.
In fact, the problem of inefficient photon-to-magnon coupling is also relevant in classical magnonic devices \cite{Csaba_PLA_381_1471}, which forces developments of alternative-to-microwave excitation/detection schemes \cite{Kajiwara_Nat_464_262}, and also stimulates the development of alternative types of magnon-magnon hybrids \cite{Klingler_PRL_120_127201,Li_PRL_124_117202}.

In this work, we demonstrate realization of ultra-strong photon-to-magnon coupling with the following peak characteristics: the coupling strength above 2~GHz, the single-spin coupling strength about a 100~Hz, the cooperativity reaching 240, and the coupling constant reaching 0.85.
The later parameter indicates that about 70\% of the total energy in the system is swapped between the photons and magnons within a single oscillation period of the individual uncoupled resonators.
These characteristics are achieved as a result of electromagnetic interaction between two subsystems: a superconductor/insulator/superconductor thin film hetero-structure, where the phase velocity of photons is substantially reduced, and a superconductor/ferromagnetic/superconductor thin film hetero-structure, where superconducting proximity at both interfaces enhances the collective spin eigen-frequencies\cite{Li_ChPL_35_077401,Jeon_PRAppl_11_014061,Golovchanskiy_SFS}.

As it turned out, by addressing the photon-to-magnon coupling problem with our system, we have stumbled upon a new manifestation of superconducting spintronics \cite{Linder_NatPhys_11_307,Eschrig_RPP_78_104501}. 
Superconducting spintronics is based on the phenomenon of superconducting proximity in superconductor/ferromagnetic systems and considers the ability of a ferromagnet to carry the superconducting condensate. 
The later is actually the challenge, owing to antagonistic ordering of spins by superconducting and ferromagnetic phenomena, and requires to build systems with rather exotic ferromagnetic materials \cite{Ryazanov_PRL_86_2427,Robinson_Sci_329_59,Larkin_APL_100_222601}, or with technologically-sophisticated ultra-thin-film nanostructures \cite{Kapran_PRR_2_013167,Parlato_JAP_127_193901}.
In this work, we demonstrate the existence of superconducting coherence within superconductor/ferromagnetic/superconductor three-layers that is accompanied by the magnetization precession.
Large thickness of the ferromagnetic layer in comparison to typical superconducting-proximity coherence length scales suggests the spin-triplet origin of the coherence.

\subsection*{Experimental results and discussion}

\begin{figure}[!ht]
\begin{center}
\includegraphics[width=0.5\columnwidth]{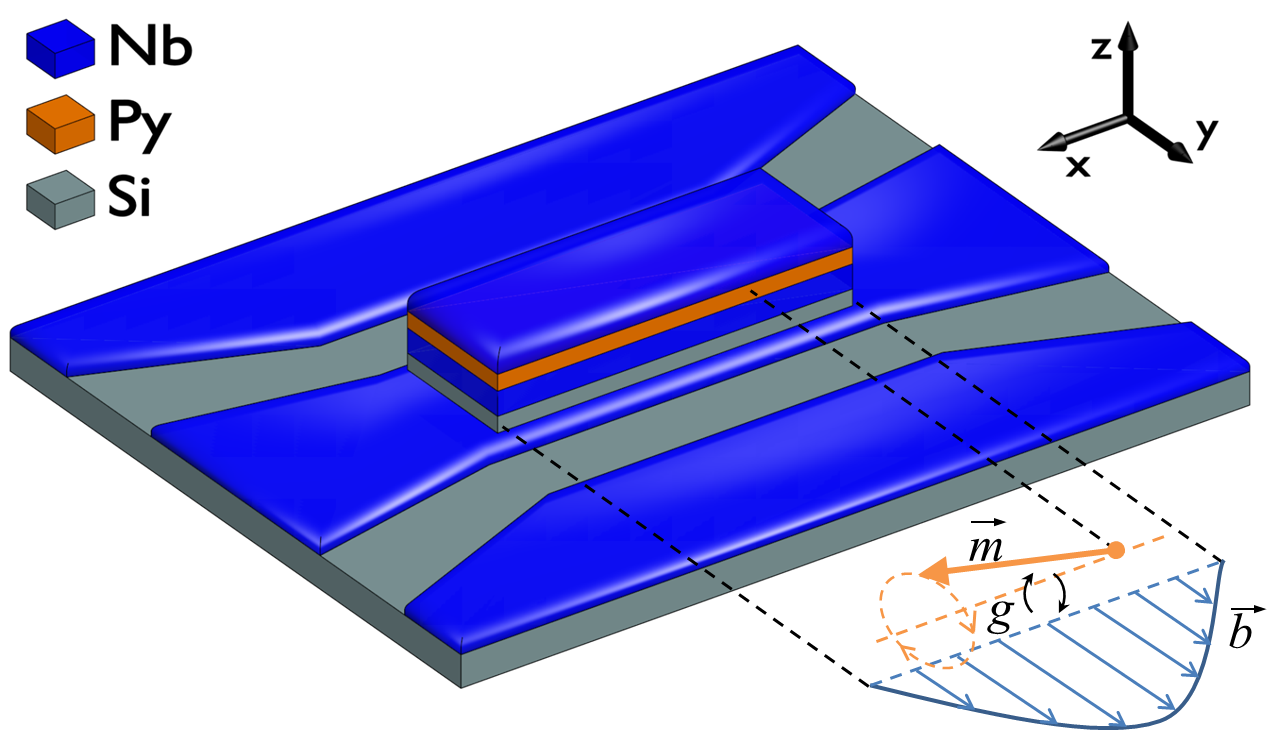}
\caption{
Schematic illustration of the investigated chip-sample.  
A series of I/S/F/S film samples of length $L=1.1$~mm along $x-$axis and width $W=130$~$\mu$m along $y-$axis is placed directly on top of the central line of superconducting co-planar waveguide (in the illustration only one such film structure is shown).
Magnetic field $H$ is applied in-plane along the $x$-axis. 
Orange arrow $\vec{m}$ indicates resonant precession of magnetization in S/F/S subsystem.
Blue curve with blue arrows $\vec{b}$ indicate magnetic field component of Swihart electromagnetic standing wave in S/I/S subsystem.
The $g$-term indicates the photon-to-magnon coupling.
}
\label{sam}
\end{center}
\end{figure}

A schematic illustration of investigated hybrid systems is shown in Fig.~\ref{sam}.
The system consists of superconducting (S) niobium (Nb) film coplanar waveguide (CPW) and multilayered rectangular film hetero-structures placed directly on the top of the transmission line.
Multilayered film heterostructures are fabricated with lateral dimensions $L\times W=1100\times130$~$\mu$m$^2$ out of Nb, ferromagnetic (F) permalloy (Py=Fe$_{20}$Ni$_{80}$) and insulating (I) Si or AlO$_x$ layers.
A number of different samples has been fabricated and measured with different thickness and order of S, F, and I layers.
The response of experimental samples was studied by analyzing the transmitted microwave signal $\left|S_{21}\right|(f,H)$ with the vector network analyzer (VNA) Rohde \& Schwarz ZVB20.
See Methods and supplementary for details.


%
\begin{figure}[!ht]
\begin{center}
\includegraphics[width=0.32\columnwidth]{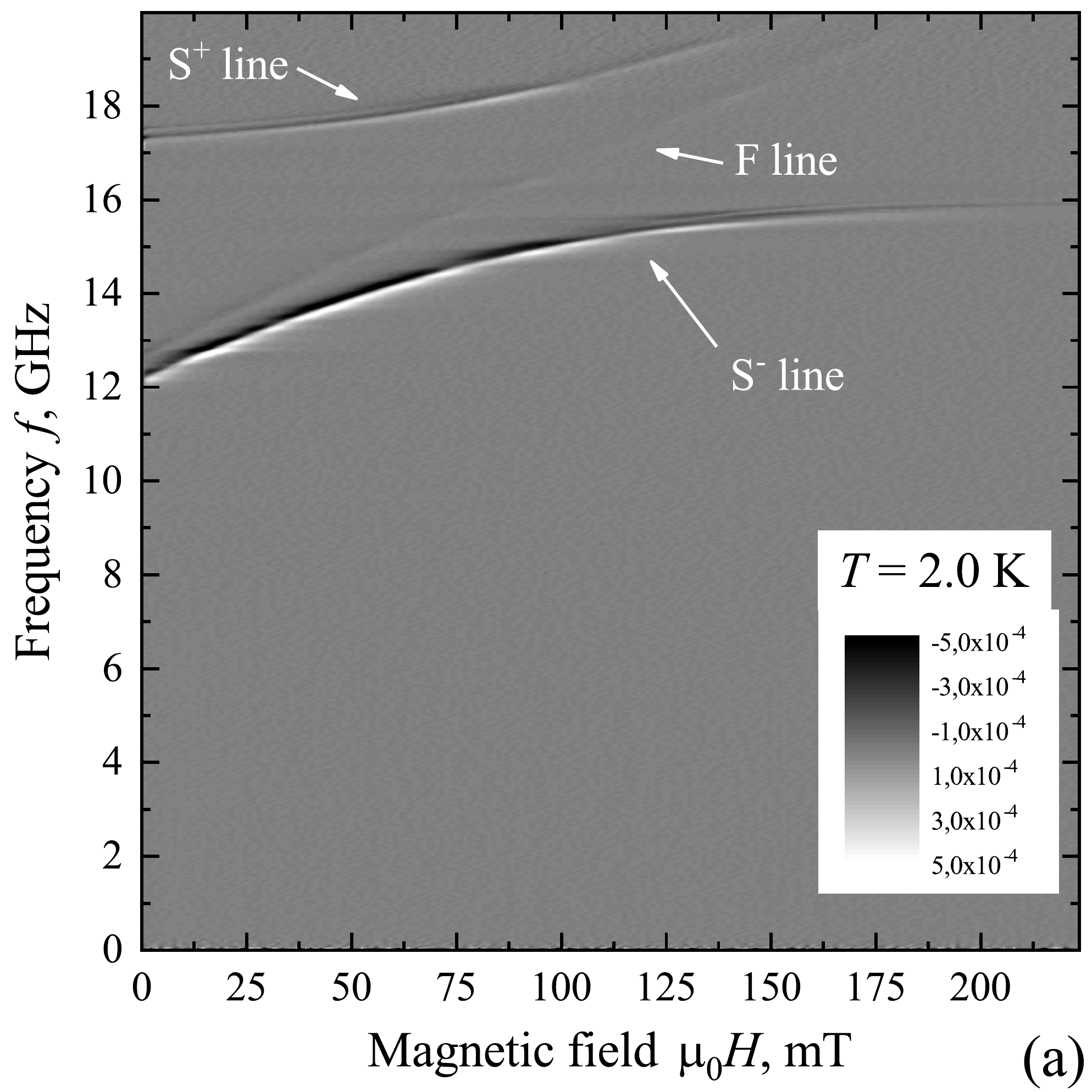}
\includegraphics[width=0.32\columnwidth]{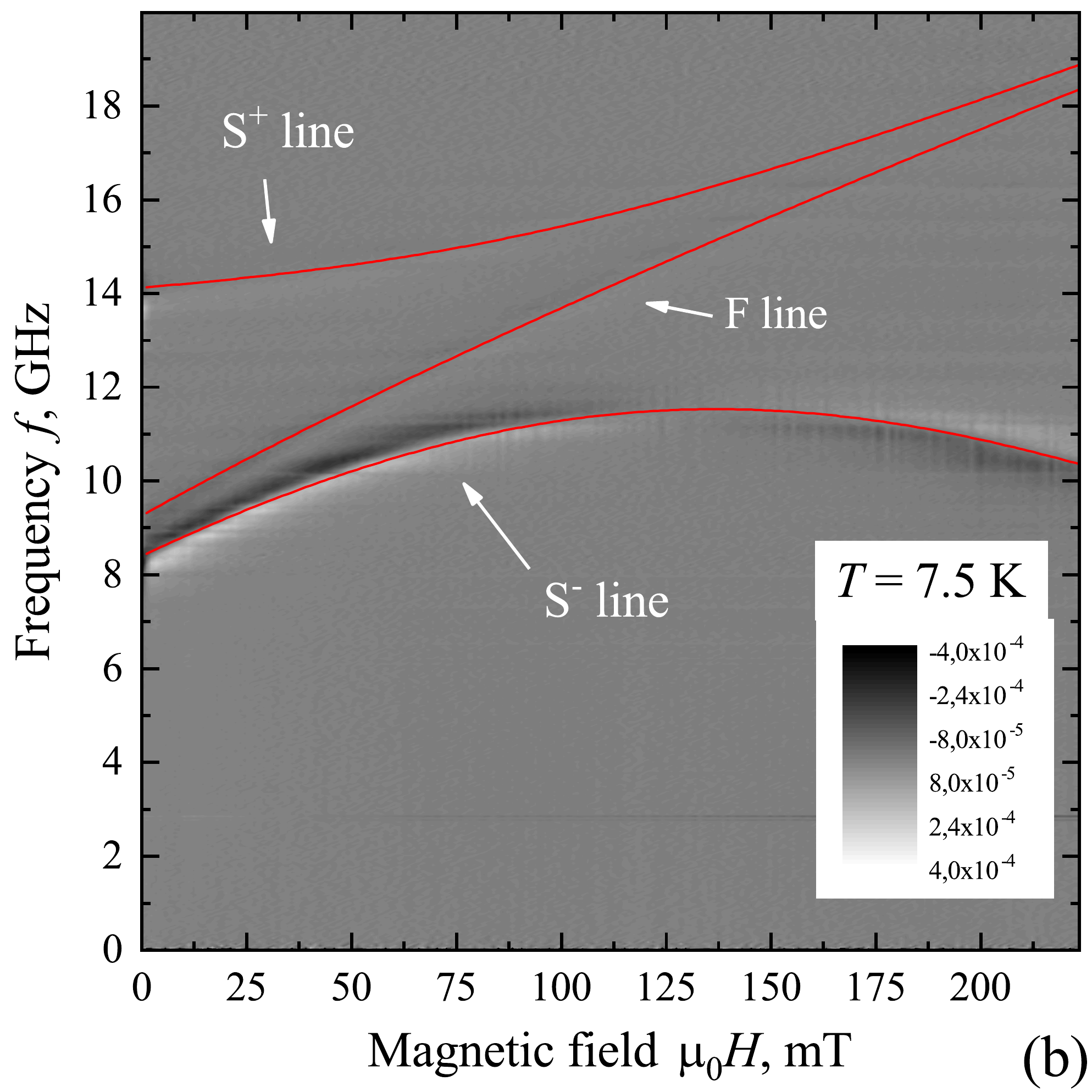}
\includegraphics[width=0.32\columnwidth]{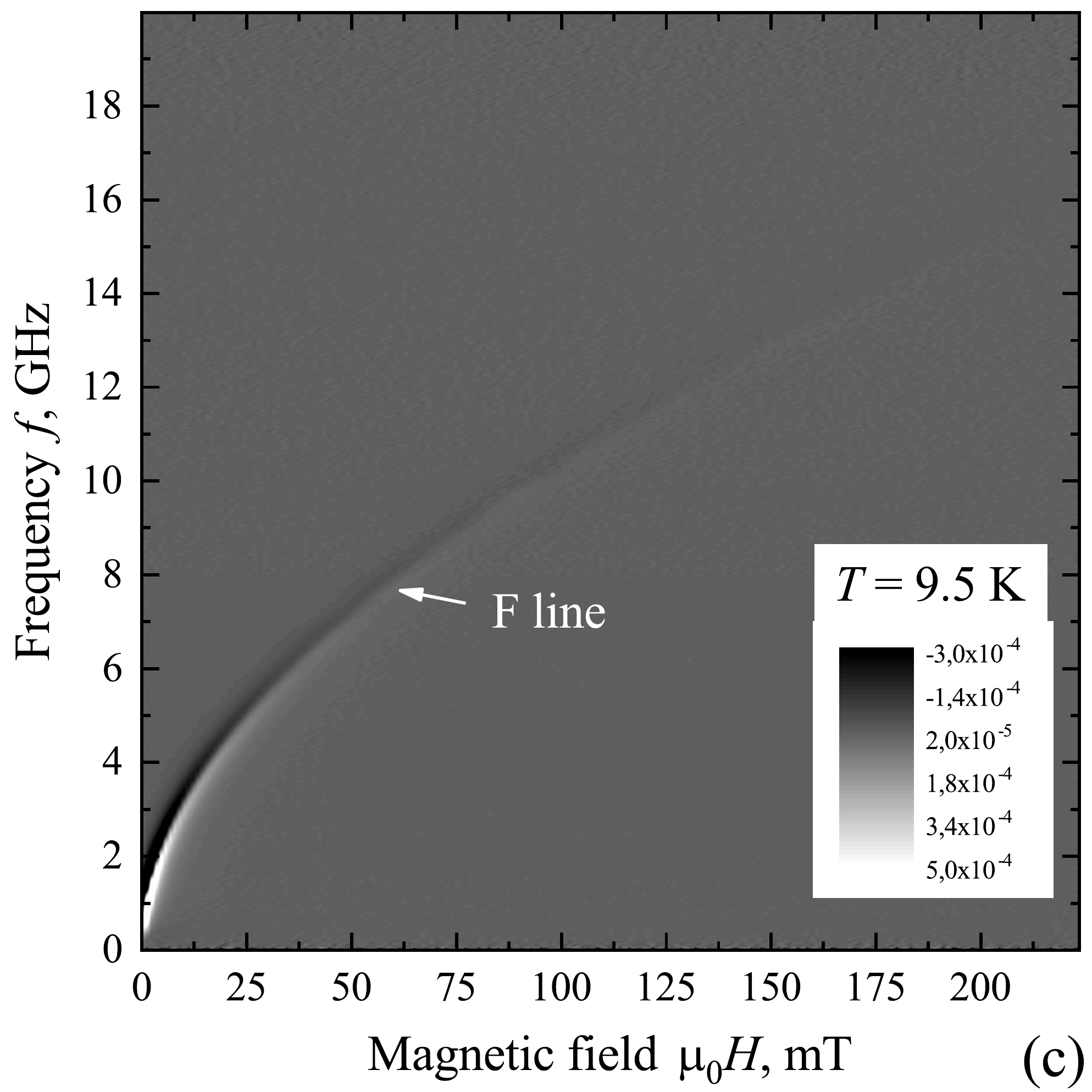}
\includegraphics[width=0.32\columnwidth]{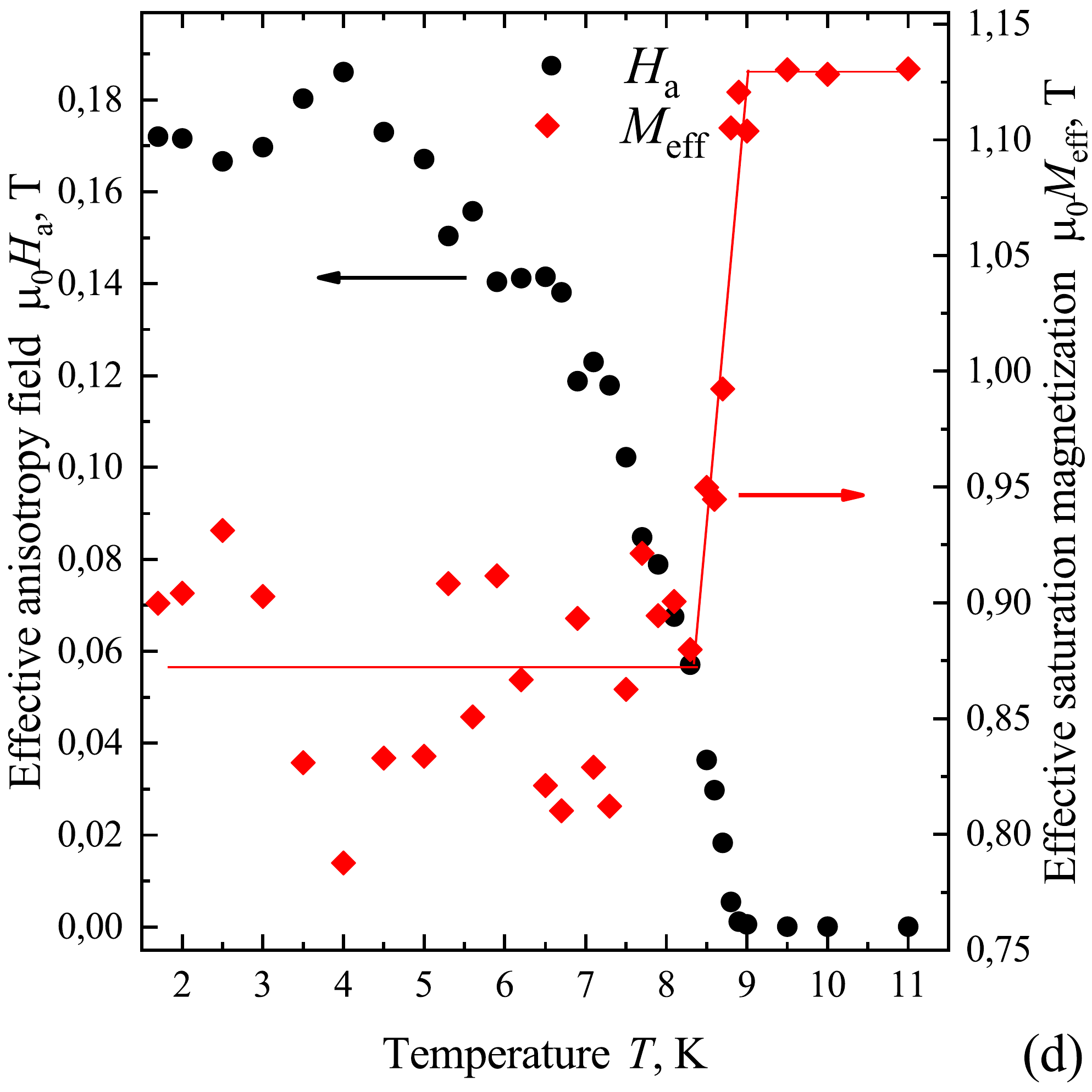}
\includegraphics[width=0.32\columnwidth]{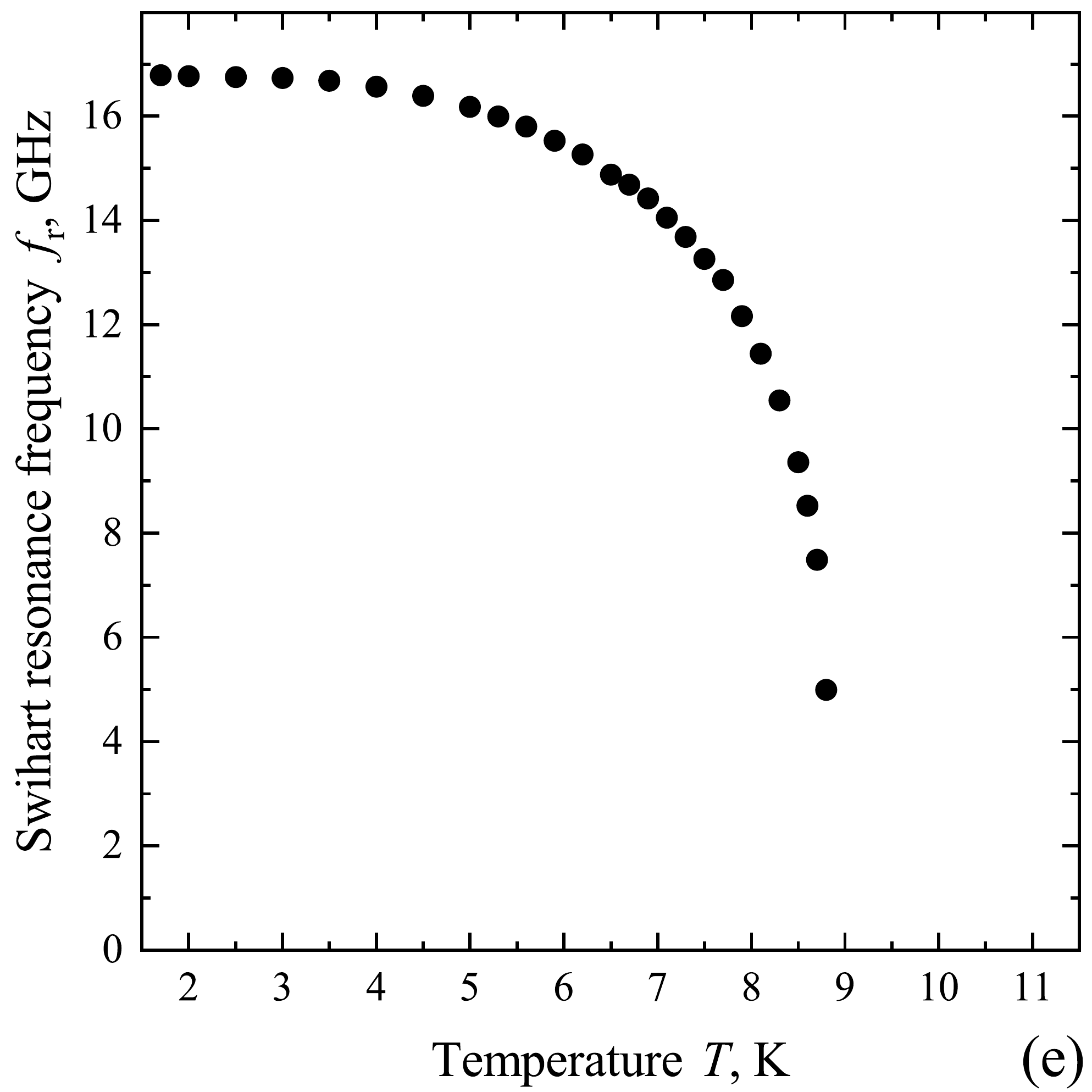}
\includegraphics[width=0.32\columnwidth]{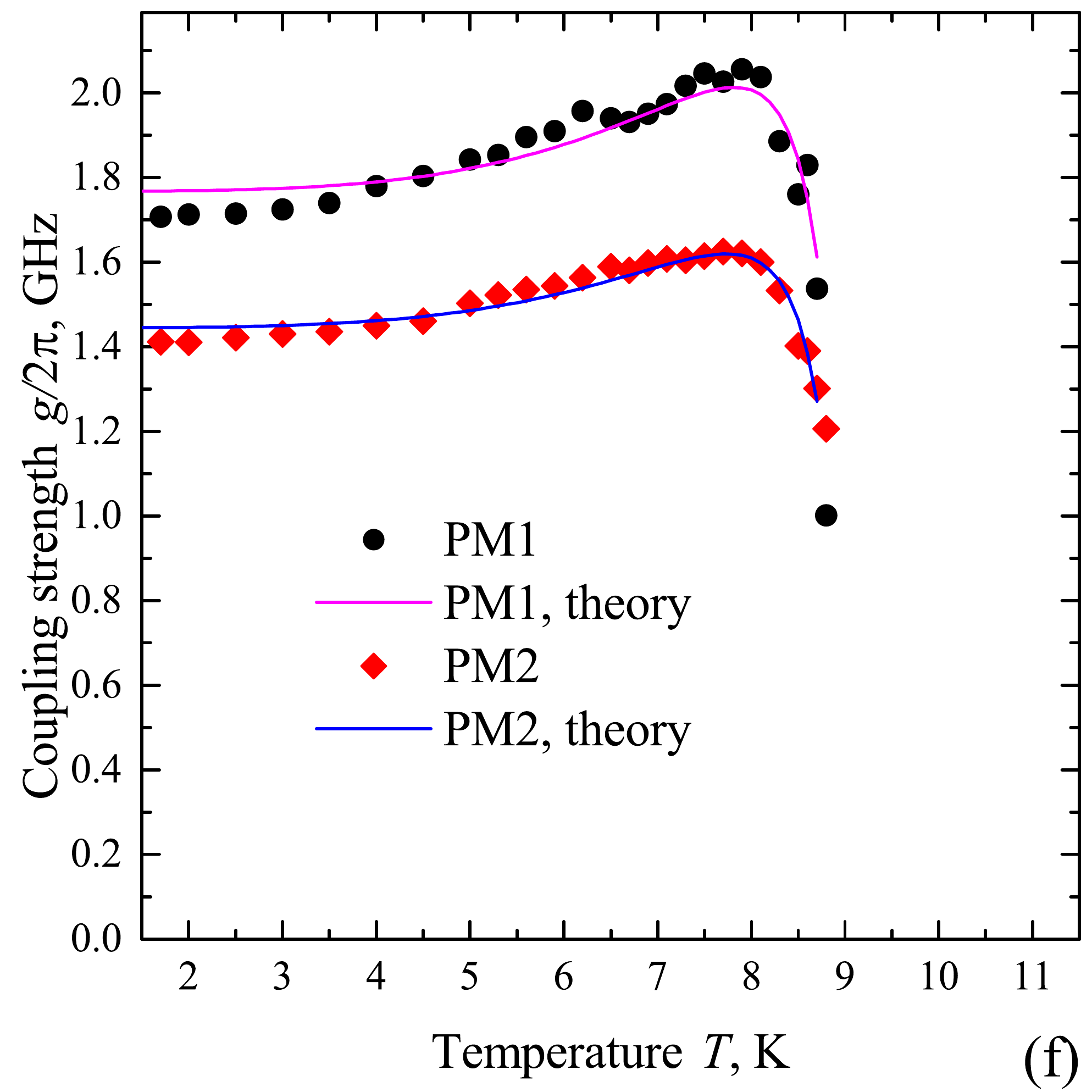}
\caption{a-c) Microwave transmission spectra $d\left|S_{21}\right|(f,H)/dH$ of the PM1 sample measured at different temperatures below (a,b) and above (c) the superconducting critical temperature.
Red curves in (b) show modeling of spectral lines with Eqs.~\ref{FMR},\ref{Swi},\ref{Coup} at corresponding temperature.
d) Temperature dependencies of the proximity-induced anisotropy $H_a(T)$ and effective magnetization $M_{eff}(T)$ of the ferromagnetic S/F/S subsystem (see Eq.~\ref{FMR}).
Red line in (d) is provided as an eye-guide for $M_{eff}(T)$.
e) Temperature dependence of the zero-field Swihart resonance frequency $f^{S0}_r(T)$ of the S/I/S subsystem (see Eq.~\ref{Swi}).
f) Temperature dependence of the coupling strength $g(T)/2\pi$ between S/F/S and S/I/S subsystems for the PM1 sample and also for the supplementary PM2 sample (see Eq.~\ref{Coup}).
Solid lines in (f) show the fit with the model Eq.~\ref{g_T}.
The optimum fit with Eq.~\ref{g_T} yields the zero-temperature London penetration depth in S/F/S multilayer $\lambda_0=78.4$~nm and $\alpha=225.9$~GHz~nm$^{3/4}$ for PM1 sample, and $\lambda_0=80.0$~nm and $\alpha=181.2$~GHz~nm$^{3/4}$ for PM2 sample.
}
\label{Exp1}
\end{center}
\end{figure}

Figure ~\ref{Exp1} collects experimental data for the PM1 sample that consist of I/S/F/S = Si(30nm)/Nb(102nm)/Py(35nm)/Nb(103nm) rectangles placed on top of 140~nm thick Nb wavequide (see Tab~\ref{Tab} and Fig.~\ref{sam}).
Figures~\ref{Exp1}a-c show spectra $d\left|S_{21}\right|(f,H)/dH$ measured at $T=2$~K (a) and $T=7.5$~K (b), which are below the superconducting temperature of Nb $T_c\approx 9$~K, and at $T=9.5$~K (c), which is slightly above the $T_c$.
At $T>T_c$ (Fig.~\ref{Exp1}c) the spectrum consist of a single absorption line indicated as the F-line.
The F-line represents the conventional ferromagnetic resonance (FMR) absorption by the F-layer.
The conventional FMR curve $f^{F}_r(H)$ for thin in-plane-magnetized ferromagnetic films at in-plane magnetic field obeys the Kittel dependence:
%
\begin{equation}
(2\pi f^{F}_r(H)/\mu_0\gamma)^2=(H+H_a)(H+H_a+M_{eff})
\label{FMR}
\end{equation}
%
where $f^{F}_r$ is the FMR frequency, $\mu_0$ is the vacuum permeability, $\gamma=1.856\times10^{11}$~Hz/T is the gyromagnetic ratio for Py, $H_a$ is the anisotropy field that is aligned with the external field, and  $M_{eff}$ is the effective magnetization.
Modeling of the absorption line at $T>T_c$ with Eq.~\ref{FMR} yields negligible $\mu_0 H_a\sim10^{-4}$~T and $\mu_0 M_{eff}=1.13$~T, which are typical for Py thin films.

At $T<T_c$ the spectrum changes drastically.
At $T=2$~K (Fig.~\ref{Exp1}a) the spectrum contains three resonance lines indicated as the F-line, S$^+$- and S$^-$-lines.
The roughly-linear dependence of the absorption on magnetic field (F-line) corresponds to the FMR absorption by the hybrid S/F/S subsystem of the PM1 sample.
As reported in Ref.~\cite{Golovchanskiy_SFS}, superconducting proximity  in S/F/S three-layers modifies anisotropy fields:
induces giant positive anisotropy $H_a$ and reduces the effective magnetization $M_{eff}$.
These changes shift the FMR to higher frequencies.

S$^+$- and S$^-$-lines in Fig.~\ref{Exp1}a are identified as the avoided crossing (a.k.a. anti-crossing or level repulsion) response \cite{Tabuchi_PRL_113_083603, Zhang_PRL_113_156401,Huebl_PRL_111_127003} of two coupled oscillators: of the ferromagnetic S/F/S resonator (F-line) and of a microwave photon resonator.
The later is characterized by the resonance frequency of about 17~GHz at zero field (frequency of the S$^+$-line at zero field).
The microwave photon resonator is formed at the insulating layer of the PM1 sample between the Nb-CPW transmission line and the first S-layer of the deposited I/S/F/S multilayer (see Fig.~\ref{sam}).
Indeed, in the insulating film constrained by two S-layers the photon phase velocity is reduced following the expression 
%
\begin{equation}
\overline{c}=c_0\sqrt{d_I / \varepsilon_I(2\lambda_L+d_I)}
\label{LV}
\end{equation}
%
where $c_0$ is the velocity of light in vacuum, $\overline{c}$ is the modified velocity of light known as the Swihart velocity, $d_I$ is the thickness of the I-layer, $\varepsilon_I$ is the dielectric constant of the I-layer, and $\lambda_L$ is the London penetration depth of the S-layer.
Considering $d_I=30$~nm of the Si layer, $\varepsilon_I\approx 10$ for Si, and $\lambda_S\approx 90$~nm in Nb one obtains $\overline{c}=0.12c_0$ in S/I/S subsystem of the PM1 sample,
which provides the resonance frequency $f^{S}_r\approx 17$~GHz for a $\lambda/2$-resonator with the length of deposited rectangles $L=1.1$~mm.
This frequency matches exactly the resonance frequency of the microwave resonator at zero field in Fig.~\ref{Exp1}a.
Therefore, S$^+$- and S$^-$-lines correspond to the response of coupled microwave S/I/S and ferromagnetic S/F/S resonators.
%

A more detailed understanding of the system can be obtained by analyzing the temperature dependence of the microwave transmission spectrum at $T<T_c$.
Figure~\ref{Exp1}b shows the spectrum of the PM1 sample at $T=7.5$~K$<T_c$.
Visual comparison of Fig.~\ref{Exp1}a and Fig.~\ref{Exp1}b shows that upon increasing temperature the F-line shifts to lower frequencies, which is consistent with temperature dependence of the FMR in S/F/S systems \cite{Golovchanskiy_SFS}.
Both S$^+$- and S$^-$-lines also shift to lower frequencies.
For instance, the zero-field frequency of the Swihart resonator shifts down to about 14~GHz (see S$^+$-line at zero field).
This temperature dependence of the Swihart resonance is provided by the temperature dependence of the London penetration depth $f^{S}_r\propto 1/\sqrt{\lambda_L}$ (see Eq.~\ref{LV}).
Also, Fig.~\ref{Exp1}b shows that the resonance frequency given by S$^-$-line decreases upon increasing magnetic field at $\mu_0 H>120$~mT.
The field dependence of the Swihart resonance frequency is provided by the field dependence of the London penetration depth of the s-wave superconductors $\lambda_L\propto H^2$ \cite{Hanaguri_PhysC_246_223}, and also by the dependence of the effective penetration depth on thickness of superconducting film \cite{Gubin_PRB_72_064503}, which transforms $\lambda_L\propto H^2$ to $\lambda_S\propto H^4$.
At this stage, the model for the Swihart resonance frequency $f^{S}_r(T,H)$ can be proposed:
%
\begin{equation}
f^{S}_r(T,H)=f^{S0}_r(T)/\sqrt{1+\alpha_1(T)H^2+\alpha_2(T)H^4}
\label{Swi}
\end{equation}
%
where $f^{S0}_r(T)$ is the temperature-dependent zero-field Swihart resonance frequency, $\alpha_1(T)$ and $\alpha_2(T)$ are free temperature-dependent parameters.

Summarizing, transmission spectra in Figs.~\ref{Exp1}a,b show the FMR absorption in S/F/S subsystem, which follows the Kittel field dependence (Eq.~\ref{FMR}), and the collective response of two harmonic oscillators indicated as S$^+$- and S$^-$-lines: of the ferromagnetic S/F/S resonator and of the Swihart S/I/S resonator (Eq.~\ref{Swi}).
The coupling between the S/F/S and S/I/S resonators is schematically depicted in Fig.~\ref{sam}.
When two harmonic oscillators are coupled their resonance spectrum is represented by the anti-crossing pattern \cite{Tabuchi_PRL_113_083603, Zhang_PRL_113_156401,Huebl_PRL_111_127003}
%
\begin{equation}
f^{+(-)}_r=\frac{f^{S}_r+f^{F}_r}{2} \pm \sqrt{\left(\frac{f^{S}_r+f^{F}_r}{2}\right)^2+\left(\frac{g}{2\pi}\right)^2}
\label{Coup}
\end{equation}
%
where $g/2\pi$ is the coupling strength.

Equations~\ref{FMR},~\ref{Swi},~and~\ref{Coup} are employed for quantitative analysis of microwave transmission spectra at different temperatures using the following routine.
First, the F-line is fitted separately at each temperature with Eq.~\ref{FMR}. 
The fit yields temperature-dependent proximity-induced anisotropy $H_a(T)$ and effective magnetization $M_{eff}(T)$ shown in Fig.~\ref{Exp1}d.
Both $H_a(T)$ and $M_{eff}(T)$ are well consistent with proximity-induced anisotropies in S/F/S systems \cite{Golovchanskiy_SFS}.
Next, S$^+$- and S$^-$-lines are fitted with Eq.~\ref{Coup} using $H_a(T)$ and $M_{eff}(T)$ as fixed parameters and using parameters of the Swihart resonator (Eq.~\ref{Swi}) and the coupling strength as fitting parameters.
Examples of modeling of resonance lines of the PM1 sample with Eqs.~\ref{FMR},~\ref{Swi},~and~\ref{Coup} are given in Fig.~\ref{Exp1}b and in supplementary.
The optimum fit yields the temperature-dependent zero-field Swihart resonance frequency $f^{S0}_r(T)$ and the coupling strength $g(T)/2\pi$ given in Figs.~\ref{Exp1}e~and~\ref{Exp1}f, respectively.
Zero-field Swihart resonance frequency $f^{S0}_r(T)$ decreases with temperature reaching zero at $T=T_c$ owing to the temperature dependence of the London penetration depth $f^{S0}_r(T)\propto 1/\sqrt{\lambda_L(T)}\propto(1-(T/T_c)^4)^{1/4}$.

The high value of the coupling strength and its dependence on temperature $g(T)/2\pi$ are the key achievements of this work.
The curve $g(T)/2\pi$ in Fig.~\ref{Exp1}f is non-monotonous: upon increasing temperature it grows progressively from about 1.7~GHz up to a peak value of 2.06~GHz at $T\approx 8$~K, and then decreases rapidly while approaching the critical temperature.
The achieved peak coupling strength is the record value among photon-to-magnon hybrids, exceeding the coupling strength in cavity-based \cite{Tabuchi_PRL_113_083603, Zhang_PRL_113_156401,Lachance-Quirion_APE_12_070101}, split-ring-based \cite{Bhoi_SciRep_7_11930}, and on-chip resonator-based \cite{Li_PRL_123_107701,Hou_PRL_123_107702} hybrids by an order of the magnitude.
Importantly, along with the total coupling the system demonstrates a strong single-spin coupling strength \cite{Huebl_PRL_111_127003,Tabuchi_PRL_113_083603, Zhang_PRL_113_156401,Bhoi_SciRep_7_11930,Lachance-Quirion_APE_12_070101} $g_s/2\pi=g/2\pi/\sqrt{N}=88$~Hz at $T=8$~K, where $N$ is the number of spins in the system.
This value of $g_s/2\pi$ exceeds ones for cavity-based \cite{Tabuchi_PRL_113_083603, Zhang_PRL_113_156401,Lachance-Quirion_APE_12_070101}, split-ring-based \cite{Bhoi_SciRep_7_11930}, and flip-chip-based \cite{Huebl_PRL_111_127003} hybrids by several orders of the magnitude and is only comparable with coupling values achieved recently in on-chip hybrids \cite{Li_PRL_123_107701,Hou_PRL_123_107702}.


All statements above are verified with a supplementary PM2 sample that consists of a similar Swihart resonator and a different-volume ferromagnetic subsystem.
Such system is expected to demonstrate different total coupling strength with the same single-spin coupling strength as compared to the PM1 sample.
Supplementary~S1 collects experimental data for the PM2 sample that consist of I/S/F/S = Si(15nm)/Nb(110nm)/Py(19nm)/Nb(110nm) rectangles placed on top of 120~nm thick Nb wavequide (see Tab~\ref{Tab} and Fig.~\ref{sam}).
%
The coupling strength $g(T)/2\pi$ for PM2 sample is given in Fig.~\ref{Exp1}f.
The curve $g(T)/2\pi$ shows similar temperature dependence with one for PM1 sample: the maximum coupling of 1.67~GHz is reached at $T\approx 8$~K.
As expected, at fixed $T$ the coupling strength in PM2 sample is reduced as compared to PM1 sample, owing to thinner F-layer, while the single-spin coupling strength of PM2 sample at $T=8$~K $g_s/2\pi=92$~Hz agrees with one for PM1 sample.

As a final remark to this subsection, we discuss two practical characteristics of coupling.
The coupling constant $k$ represents the information exchange rate between the magnon and photon modes during interaction at specified frequency \cite{Bhoi_SciRep_7_11930} and is defined as
$ k=\sqrt{2(g/2\pi)/f^{S}_r} $.
Estimations show that the coupling constant for PM1 and PM2 samples reaches 0.85, which indicate the ultra-strong coupling regime with record strength among hybrid magnonic systems (see Refs.~\cite{Zhang_PRL_113_156401,Flower_NJP_21_095004} for comparison).
%
Another important practical parameter is the cooperativity $C$ \cite{Huebl_PRL_111_127003,Li_PRL_123_107701} that is defined as 
$ C=(g/2\pi)^2/\Delta f^{S^+}_r \Delta f^{S^-}_r $,
where $\Delta f^{S^+}_r$ and $\Delta f^{S^-}_r$ are the linewidth of the $S^+$- and $S^-$-lines, respectively, taken at magnetic field of the coupling.
Cooperativity characterizes coherence between damped oscillators.
Insufficient cooperativity leads to suppression of coherent information exchange between oscillators and to damping of the signal instead of resolved avoided crossing spectrum.
Estimations provide the maximum $C=240$ for PM1 sample and $C=109$ for PM2 sample at 2~K.
These are record values for magnonic hybrids based on metallic ferromagnets, offering a great flexibility in circuit integration.
See supplementary~S3,~S6 for more details.


The fundamental reason for strong photon-to-magnon coupling in studied samples is rather straightforward.
The single-spin coupling strength is inversely proportional to the mode volume $V_c$ of electromagnetic resonator \cite{Zhang_PRL_113_156401,Huebl_PRL_111_127003,Li_PRL_123_107701} $g_s\propto 1/\sqrt{V_c}$.
The suppressed photon velocity $\overline{c}$ in Swihart resonator  (Eq.~\ref{LV}) provides a reduced resonance length and small overall dimensions of the resonator.
For instance, both the electric mode volume of the PM1 resonator $V^{SE}_c=L\times W\times d_I=4.3\times10^{-15}$~m$^3$ and its magnetic volume $V^{SM}_c\approx L\times W\times (2\lambda_L+d_I)=2.8\times10^{-14}$~m$^3$ are well comparable with the volume of the ferromagnetic layer $V^{F}=L\times W\times d_F=5.0\times10^{-15}$~m$^3$.
Using other terms, high single-spin coupling strength is provided by essentially low impedance $Z$ of the Swihart resonator \cite{Hou_PRL_123_107702} $g_s\propto 1/\sqrt{Z}$.
The impedance of the S/I/S resonator of PM1 sample at 2 K can be estimated as $Z=1/(2\pi f^S_r C)=0.02$~$\Omega$, where $C=\varepsilon_0\varepsilon LW/d_I=4.7\times10^{-10}$~F is its capacitance.

%
\begin{figure}[!ht]
\begin{center}
\includegraphics[width=0.49\columnwidth]{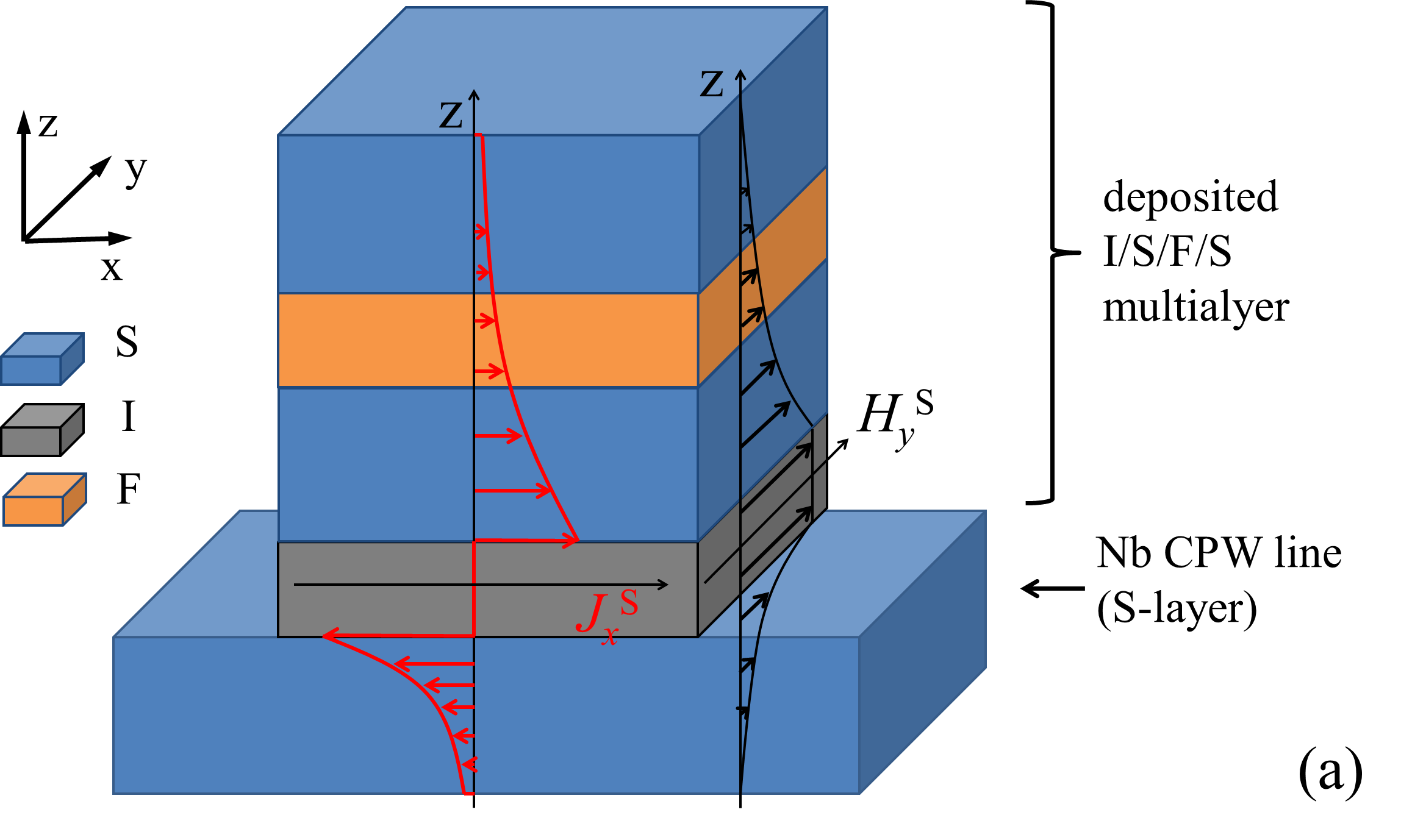}
\includegraphics[width=0.20\columnwidth]{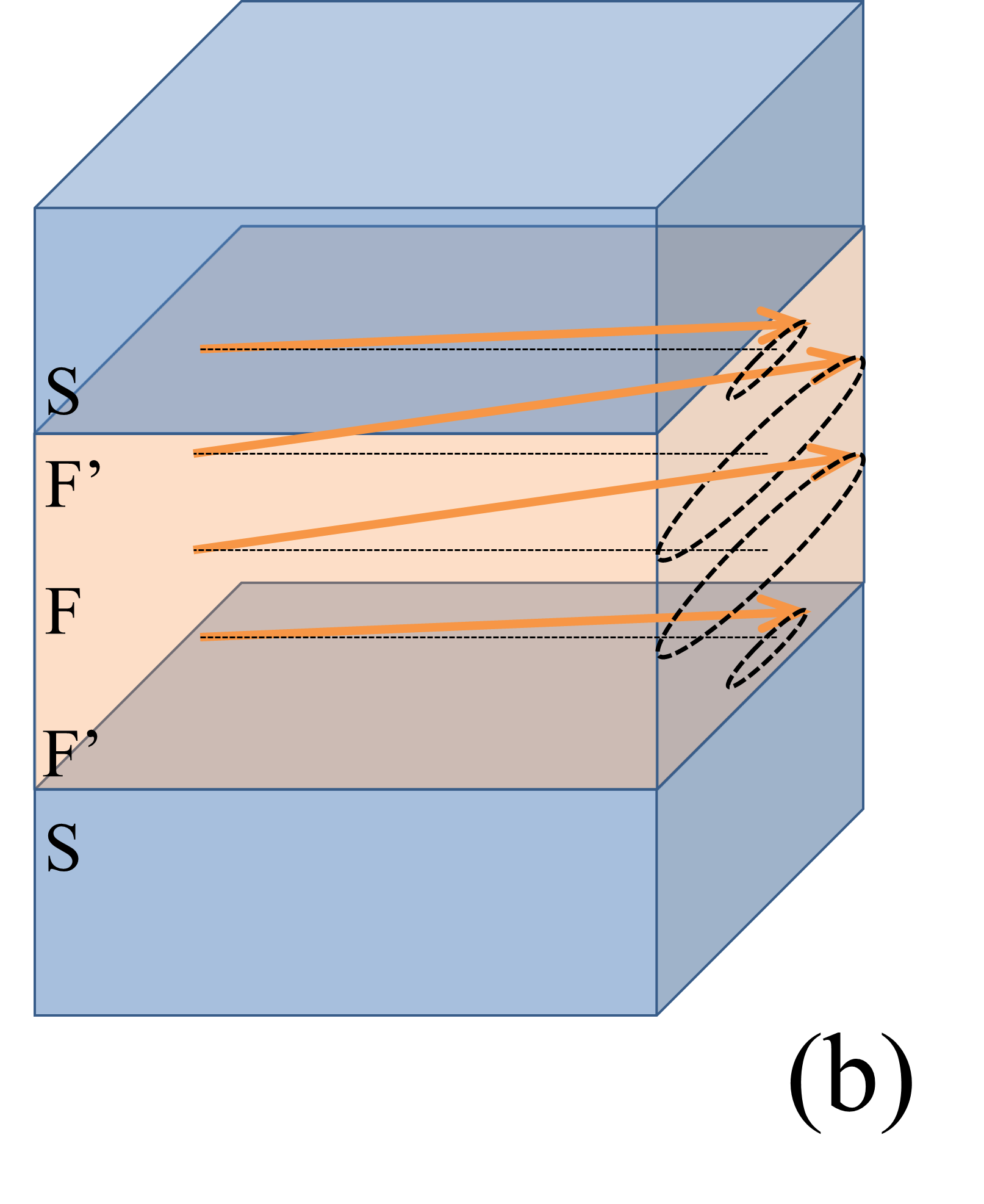}
\caption{
a) Schematic illustration of field and current distributions that lead to photon-to-magnon coupling.
Black curves and arrows indicate the distribution of the magnetic field component $H^S_y(z)$ of the electromagnetic standing wave in superconducting layers of the Swihart resonator.
Red arrows with red curves indicate the distribution of superconducting currents $J^S_x(z)$ that support the electromagnetic standing wave in superconducting layers of the Swihart resonator. 
b) Schematic illustration of the concept of dynamic spin-triplet generator.
The amplitude of magnetization precession (orange arrows and dashed circles) is suppressed in vicinity to the S/F interface as compared to the bulk of the film.
Distribution in the amplitudes forms the dynamic non-collinear F'FF' spin state. 
}
\label{Fit}
\end{center}
\end{figure}

Understanding of microscopic mechanism of the coupling is challenging and required additional experiments provided in supplementary~S4.
First, one should note that the geometry of the studied structure is essentially the infinite thin film geometry.
The presence of the photon-to-magnon coupling between S/I/S and S/F/S oscillators is equivalent to presence of nonzero alternating magnetic fields $H^S_y$ induced by the microwave resonator in the F-layer (see Fig.~\ref{Fit}).
Supplementary experiments evidence that the realization of the coupling requires both S-layers for the S/F/S subsystem, presence of the superconducting proximity at both S/F interfaces, and excludes conventional magnetostatic as a possible source of the coupling.
Combination of these factors imply that the only possibility for the coupling is when the entire S/F/S three-layer acts as a single upper superconducting layer of the Swihart resonator. 
In other words, presence of nonzero $H^S_y$ in the F-layer is equivalent to presence of coherent Meissner-like fields and superconducting currents $J^S_x$ in the entire S/F/S three-layer including the F-layer as illustrated in Fig.~\ref{Fit}.

The suggested mechanism of the coupling obeys the following quantitative model (see Supplementary~S5):
%
\begin{equation}
g(T)/2\pi=\alpha\lambda_{eff}(T)^{-3/4}\sinh{\left(\frac{d_{S}+d_F/2}{\lambda_{eff}(T)}\right)}/\sinh{\left(\frac{2d_{S}+d_F}{\lambda_{eff}(T)}\right)}
\label{g_T}
\end{equation}
%
where $\alpha$ is a free fitting parameter, 
and $\lambda_{eff}(T)$ is the effective penetration depth that depends on the London penetration depth at zero temperature $\lambda_0$, on $T_c$, on thicknesses $d_{S}$, $d_F$ in S/F/S multilayer, and on temperature $T$.
Figure~\ref{Exp1}f shows experimental and model dependencies of the coupling strength on temperature where the critical temperature $T_c=9.05$~K and the cut-off temperature 8.7~K were used.
The optimum fit with Eq.~\ref{g_T} yields $\lambda_0=78.4$~nm, for PM1 sample, and $\lambda_0=80.0$~nm for PM2 sample.
The obtained $\lambda_0$ are perfectly consistent with typical values for sputtered niobium thin films, which verifies the proposed model.
According to the model, the increase in coupling strength upon increasing temperature at $T<8$~K occurs due to the increase of $H_y^S$ in the F-layer (the sinh-term), while the rapid drop of the coupling at $T>8$~K occurs due to simultaneous increase of the mode volume of the resonator and reduction of the resonance frequency (the $\lambda_{eff}^{-3/4}$ term).


Notice that earlier research studies \cite{Li_ChPL_35_077401,Jeon_PRAppl_11_014061,Golovchanskiy_SFS} have shown that presence of superconducting layers at both sides of the F film and of superconducting proximity at both S/F interfaces of an S/F/S three-layer leads to a substantial increase in the ferromagnetic resonance frequency.
This phenomenon is also confirmed in the present study.
As a possible origin of that phenomenon it was proposed \cite{Li_ChPL_35_077401} that the long-range spin-triplet superconducting condensate is formed in the F-layer and that spin-polarized spin-triplet Cooper pairs induce additional torque on ferromagnetic moments via the spit-transfer-torque mechanism, which causes the increase in FMR frequency $f^F_r$.
Moreover, the effect of the superconducting triplet pairing on ferromagnetic anisotropies in S/F structures was recently addressed in Refs.~\cite{Johnsen_PRB_99_134516,Ruano_arXiv}.
The current study evidences presence of superconducting current in the F-layer in exactly the same conditions.
Large thickness of considered F-films 20~nm and 35~nm in comparison with the spin-singlet coherence length $\xi_F\approx 1$~nm hint for spin-triplet origin of registered superconducting coherence.  
Thus, a set of different research studies point towards formation of spin-triplet superconducting condensate in the F-layer of the S/F/S three-layers.

In general, only two mechanisms are known for formation of the spin-triplet superconductivity in S/F/S three-layers.
The first one requires presence of strong spin-orbit coupling at S/F interfaces \cite{Eschrig_RPP_78_104501}, which can hardly be related to the current study.
The second one requires presence of non-collinear ferromagnetic inhomogeneities F' at both S/F interfaces as compared to the orientation in F bulk, i.e., the so-called spin-triplet generator S/F'FF'/S structure \cite{Houzet_PRB_76_060504,Khaire_PRL_104_137002,Robinson_Sci_329_59,Eschrig_RPP_78_104501}.
We presume that the later mechanism can be responsible for formation of the spin-triplet Cooper pairs in our S/F/S induced by magnetization precession. 
Indeed, in vicinity to S/F interfaces ferromagnetic state is different from the bulk.
This state is commonly attributed to additional surface anisotropies induced by surface tensions and by chemical composition at the S/F interface, 
to enhanced Gilbert damping due to the interface roughness, 
and to difference in demagnetizing fields as compared to the bulk of the film.
At FMR any of these effects leads to reduction in amplitude of magnetization precession for spins at the interface in comparison to the bulk, i.e., to formation of dynamic non-collinear spin-state, as illustrated in Fig.~\ref{Fit}b.

\subsection*{Conclusion}

Summarizing, in this work we demonstrate a new platform for realization of the ultra-strong photon-to-magnon coupling in on-chip thin film hetero-structures between superconductor/ insulator/ superconductor electromagnetic resonator and superconductor/ ferromagnet/ superconductor ferromagnetic resonator.
High characteristics of coupling are achieved owing to suppressed photon phase velocity in electromagnetic subsystem.
The route for further enhancement of the coupling strength is rather straightforward: 
one can consider a microwave resonator with even smaller phase velocity that is fabricated using superconducting materials with larger $\lambda_L$ and dielectric materials with higher $\varepsilon$, paving the way towards deep-strong coupled systems.
In addition, magnetic materials with lower losses, including YIG or Co$_{0.25}$Fe$_{0.75}$, will further enhance the cooperativity.
Apart from quantum magnonics, the demonstrated platform offers further developments in superconducting Josephson junction based magnonic systems.

The microscopic mechanism behind the demonstrated coupling evidences excitation of superconducting coherence in superconductor/ferromagnet/superconductor three-layers via strong thick ferromagnetic layers in presence of magnetization dynamics.
The length-scales suggest spin-triplet origin of superconducting coherence opening new perspectives in microwave superconducting spintronics. 

\subsection*{Materials and Methods}

A schematic illustration of investigated hybrid systems is shown in Fig.~\ref{sam}.
The system consists of superconducting film coplanar waveguide (CPW) and multilayered rectangular film heterostructures placed directly on the  top of the transmission line.
The waveguide is fabricated out of superconducting niobium (Nb) film deposited on top of Si/SiO$_x$ substrate using magnetron sputtering of Nb, optical lithography and plasma-chemical etching techniques; the waveguide  has 50~Ohm impedance and 82-150-82~$\mu$m center-gap-center dimensions.
Multilayered film heterostructures are fabricated with lateral dimensions $L\times W=1100\times130$~$\mu$m$^2$ out of superconducting niobium (Nb), ferromagnetic permalloy (Py=Fe$_{20}$Ni$_{80}$) and insulating Si or AlO$_x$ layers using optical lithography, magnetron sputtering and the lift-off techniques.
Importantly, deposition of these layers is performed in a single vacuum cycle ensuring an electron-transparency at all-metallic Nb/Py interfaces.
Multilayered heterostructures are fabricated as a series array along CPW for enhancement of total microwave response.
A number of different samples has been fabricated and measured with different thickness and order of superconducting (S), ferromagnetic (F), and insulating (I) layers (see Tab.~\ref{Tab}).
The main text is focused on the PM1 sample

\begin{table}[h]
\begin{center}
\begin{tabular}{|c|c|c|c|c|c|c|}
\hline
Sample \# & Nb-CPW (S$_1$) & I(Si or AlO$_x$) & S$_2$(Nb) & F(Py) & I(AlO$_x$) & S$_3$(Nb) \\
\hline
PM1 & 140 & Si-30 & 102 & 35 & 0 & 103  \\
\hline
PM2 & 120 & Si-15 & 110 & 20 & 0 & 110 \\
\hline
PM3 & 140 & Si-15 & 110 & 20 & 0 & 5\\
\hline
PM4 & 140 & Si-15 & 110 & 20 & 0 & 140\\
\hline
PM5 & 500 & AlO$_x$-15 & 110 & 25 & 100 & 110\\
\hline
\end{tabular}
\caption{IDs and thicknesses of layers in studied samples given in nm.}
\label{Tab}
\end{center}
\end{table}

The experimental chip was installed in a brass sample holder and wire bonded to printed circuit board with RF connectors. 
A thermometer and a heater were attached directly to the holder for precise temperature control.
The holder was placed in a superconducting solenoid inside a dry closed-cycle He4 cryostat (Oxford Instruments Triton).
The response of experimental samples was studied by analyzing the transmitted microwave signal $\left|S_{21}\right|(f,H)$ with the vector network analyzer (VNA) Rohde \& Schwarz ZVB20.
For removal of background parasitic resonance modes from consideration, all measured spectra $\left|S_{21}\right|(f,H)$ have been normalized with $\left|S_{21}\right|(f)$ at $\mu_0H=0.3$~T, and differentiated numerically in respect to $H$.
The response of experimental samples was studied in the field range from -0.22 T to 0.22 T, in the frequency range from 0 up to 20 GHz, and in the temperature range from 1.7 to 11 K.

\subsection*{Acknowledgments}

The authors acknowledge Dr. M. Silaev for fruitful discussions.
This work was supported by the Ministry of Science and Higher Education of the Russian Federation, by the Russian Science Foundation, and the Russian Foundation for Basic Research.



\subsection*{Competing interests}

The authors declare no competing interests.

\subsection*{References}

\bibliography{A_Bib_SISFS}

\end{document}